\documentclass[aps,prb,amsmath,amssymb,twocolumn,floatfix]{revtex4-1}
\usepackage{graphicx}
\usepackage{dcolumn}
\usepackage{bm}

\begin{document}

\title{Absence of $\mu$SR evidence for magnetic order in the pseudogap phase of Bi$_{2+x}$Sr$_{2-x}$CaCu$_2$O$_{8+\delta}$}
\author{S.~Gheidi,$^1$ K.~Akintola,$^1$ A. C. Y.~Fang,$^1$ Shyam Sundar,$^1$ A. M.~C\^{o}t\'{e},$^{1,2}$ S. R.~Dunsiger,$^{1,3}$ G. D.~Gu,$^4$ and J. E.~Sonier$^1$}

\affiliation{$^1$Department of Physics, Simon Fraser University, Burnaby, British Columbia V5A 1S6, Canada \\
$^2$Kwantlen Polytechnic University, Richmond, British Columbia V6X 3X7, Canada \\
$^3$Centre for Molecular and Materials Science, TRIUMF, Vancouver, British Columbia V6T 2A3, Canada \\
$^4$Brookhaven National Laboratory, Upton, New York, 11973, USA}  

\date{\today}
\begin{abstract}
We present an extended zero-field muon spin relaxation (ZF-$\mu$SR) study of 
overdoped Bi$_{2+x}$Sr$_{2-x}$CaCu$_2$O$_{8+\delta}$ (Bi2212) single crystals, intended to elucidate the origin of weak quasistatic magnetism 
previously detected by $\mu$SR in the superconducting and normal states of optimally-doped and overdoped samples. New results on heavily-overdoped single crystals show a 
similar monotonically decreasing ZF-$\mu$SR relaxation rate with increasing temperature that persists above the pseudogap (PG) temperature $T^*$ and does not evolve 
with hole doping ($p$). Additional measurements using an ultra-low background apparatus confirm that this behavior is an intrinsic property of Bi2212, which 
cannot be due to magnetic order associated with the PG phase. Instead we show that the temperature-dependent relaxation rate is most likely
caused by structural changes that modify the contribution of the nuclear dipole fields to the ZF-$\mu$SR signal. Our 
results for Bi2212 emphasize the importance of not assuming the nuclear-dipole field contribution is independent of temperature in ZF-$\mu$SR studies of
high-temperature (high-$T_c$) cuprate superconductors, and do not support a recent $\mu$SR study of YBa$_2$Cu$_3$O$_{6+x}$ that claims to 
detect magnetic order in the PG phase.    
\end{abstract}
\maketitle
\section{Introduction}
The origin of the pseudogap (PG) phase in cuprate superconductors is an enduring mystery that is widely believed to hold the key to understanding 
high-$T_c$ superconductivity. Polarized neutron diffraction (PND) measurements have produced evidence for the existence of an unusual intra-unit cell (IUC) magnetic 
order in the PG region of several cuprate families\cite{Fauque:06,Mook:08,Li:08,Baledent:10,Baledent:11,Li:11,Almeida:12,Mangin:14,Mangin:15,Mangin:17,Tang:18}, which has given credence 
to an orbital loop-current model for the PG.\cite{Varma:97,Varma:06,Varma:14}
Yet this claim remains controversial and has been challenged by an independent PND study of underdoped YBa$_2$Cu$_3$O$_{6+x}$ (Y123), 
which found no evidence for orbital magnetic order in the PG phase.\cite{Croft:17} The ensuing debate\cite{Bourges:18,Croft:18} has 
not quelled the controversy over the PND results, emphasizing the need to confirm the magnetic order by other experimental methods.
 
The IUC magnetic order detected by PND has an ordered moment of $\sim$0.1 $\mu_\mathrm{B}$ and 
should be detectable by the local probe techniques muon-spin relaxation ($\mu$SR) and nuclear magnetic resonance (NMR), unless it is rapidly fluctuating beyond the dynamical range of these methods. \cite{Mangin:14}
To date, no evidence for magnetic order in the PG phase has been found by NMR.\cite{Strassle:08,Strassle:10,Mounce:13,Wu:15}    
While an initial ZF-$\mu$SR study of Y123 reported the onset of magnetism near $T^*$,\cite{Sonier:01} it was subsequently shown that the results
can be explained by the effects of charge order in the CuO chains on the nuclear dipole fields sensed by the positive muon, as well as the onset of
muon diffusion above $T \! \sim \! 150$~K.\cite{Sonier:02} This understanding of the ZF-$\mu$SR results on Y123 was reinforced in a
comparative ZF-$\mu$SR investigation of small high-quality single crystals grown at the University of British Columbia (UBC) and a large underdoped single crystal
in which IUC magnetic order was reportedly detected by PND.\cite{Sonier:09} Only in the latter sample was any evidence for IUC magnetic order 
found, but only in 3~\% of the sample and hence apparently associated with an impurity phase.
A subsequent $\mu$SR investigation of an underdoped ($p \! = \! 0.11$) Y123 UBC sample found no evidence for magnetic order down to 0.025~K,
and placed a lower limit of $10^9$~Hz on the fluctuation rate of any potential PG magnetic order.\cite{Pal:16} 

More recently, Zhang {\it et al.} have claimed to detect the onset of slow magnetic fluctuations in Y123 at $T^*$ by $\mu$SR.\cite{Zhang:18}
Yet neither the effects of the CuO chains or muon diffusion on the nuclear-dipole contribution to the $\mu$SR signal were disentangled from
the measurements. Because hole doping in Y123 is dependent upon the oxygen content and order in the CuO chain layer, the effects of the CuO 
chains on the $\mu$SR signal are expected to change with increased hole doping. In addition to the influence of charge order on the nuclear dipole 
contribution, magnetic correlations are induced in the chain layer by oxygen vacancies,\cite{Chen:09} and both the charge and magnetic chain correlations 
are imprinted on the CuO$_2$ plane.

If the $\mu$SR signal in Y123 is a signature of slow magnetic fluctuations associated with magnetic order in the PG phase, similar results are
expected in $\mu$SR investigations of cuprate superconductors without CuO chains. Yet no evidence of PG magnetic order of any kind
has been detected in La$_{2-x}$Sr$_x$CuO$_4$ (La214) by $\mu$SR.\cite{MacDougall:08,Huang:12,Sonier:13} Instead, the ZF-$\mu$SR relaxation rate in 
La214 measured over a wide hole-doping range ($0.13 \! \leq \! p \! \leq \! 0.30$) evolves little with $p$ and is essentially independent of
temperature up to $T \! \sim \! 150$~K --- above which a significant decrease in the relaxation rate signifies the onset of muon diffusion.

The situation for Bi2212 is less clear.
A low-temperature ($T \! \leq \! 5$~K) $\mu$SR investigation of optimally-doped ($p \! = \! 0.16$) and underdoped ($p \! = \! 0.094$) Bi2212 single crystals 
did not find evidence of magnetic order down to 0.024~K.\cite{Pal:16} However, a subsequent $\mu$SR study of optimally-doped ($p \! = \! 0.16$, $T_c \! = \! 91$~K) 
and overdoped ($p \! = \! 0.198$, $T_c \! = \! 80$~K) Bi2212 single crystals (which we henceforth refer to as OP91 and OD80) at temperatures up to 200~K, 
provided evidence for weak quasistatic magnetic fields of apparently electronic origin in the superconducting and PG phases.\cite{Pal:18} 
Due to the onset of muon diffusion above $\sim \! 160$~K (similar to Y123 and La214), 
it was not possible to determine whether the onset of the magnetic fields coincide with $T^*$. Moreover, because the size of the corresponding field sensed 
by the muon is only on the order of 1~G, it is unclear whether the quasistatic magnetic fields stem from a magnetically-ordered phase.
    
To overcome the limitations imposed by muon diffusion, we have carried out additional ZF-$\mu$SR measurements on heavily-overdoped Bi2212 single crystals 
with expected values of $T^*$ below the onset temperature for muon diffusion. Using a specialized background suppression setup, we demonstrate that
the previously identified quasistatic magnetic fields definitely originate from the Bi2212 samples, but are not associated with the PG phase.
Instead we show that the variation of the corresponding ZF-$\mu$SR relaxation rate with temperature can be explained by a change in the nuclear dipole fields 
sensed by the muon due to changes in the crystal structure. 

\begin{figure}
\centering
\includegraphics[width=\columnwidth]{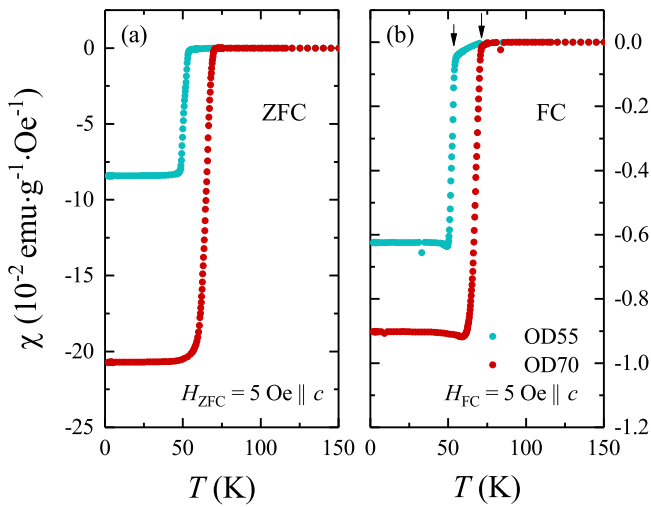}
\caption{Temperature dependence of the bulk magnetic susceptibility for the OD55 (blue circles) and OD70 (red circles) samples measured under 
(a) zero-field cooled (ZFC) and (b) field-cooled (FC) conditions. The slight decrease of the OD55 data with decreasing temperature between 
the two downward pointing arrows in (b) indicates the onset of superconductivity in some small volume fraction of the OD55 sample prior to the 
main superconducting transition at 55~K.}
\label{fig1}
\end{figure}

\section{Experimental}
Single crystals of Bi2212 were grown by the traveling-solvent-floating-zone method \cite{Hobou:09} at Brookhaven National Laboratory. The doping level 
was adjusted by tuning the excess oxygen content. In particular, overdoped single crystals were attained by annealing in an oxygen partial pressure of 
2.3~atm for 72-250 hours at 400~$^{\circ}$C. The superconducting critical temperature $T_c$ for the new samples estimated from bulk magnetic susceptibility 
measurements (see Fig.~\ref{fig1}) are $T_c \! = \! 70$~K and 55~K, corresponding to hole-doping concentrations $p \! \sim \! 0.213$ and $p \! \sim \! 0.23$, respectively. 
Henceforth, we refer to these samples as OD70 and OD55. The OD55 sample is a 0.117~g single crystal that has an $ab$-plane surface area of $\sim \! 0.68$~cm$^2$.
The OD70 sample is a thinner 0.0259~g single crystal with an $ab$-plane surface area of $\sim \! 0.25$~cm$^2$. 
Raman scattering measurements indicate that the PG phase vanishes at $p \! \sim \ 0.22$,\cite{Benhabib:15} and
hence should be absent in the OD55 sample. However, we note that field-cooled bulk magnetic susceptibility data for the OD55 sample shows an initial decrease 
below $T \! \sim \! 70$~K [see Fig.~\ref{fig1}(b)], indicating there are some phase-separated superconducting regions of somewhat higher $T_c$ in the bulk 
or at the surface. 

\begin{figure}
\centering
\includegraphics[width=\columnwidth]{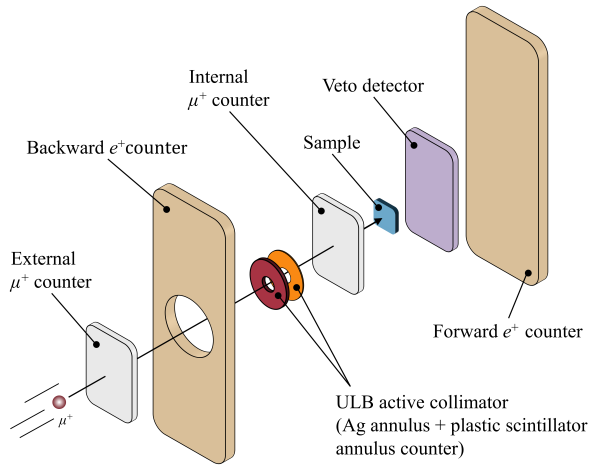}
\caption{Schematic of the arrangement of the external muon, external positron, internal muon and internal veto counters, where external and internal refer
to outside and inside the helium-gas flow cryostat, respectively. Muons passing through the internal $\mu^+$ counter that trigger the veto counter 
are rejected. Also shown are internal Ag and plastic scintillator counter annuli with equivalent 3~mm-diameter holes.  
These serve as a second type of veto (``active collimator'') in the ``ultra-low background'' (ULB) setup. Muons stopping in the Ag annulus do not pass through 
the internal muon counter and are rejected. The Ag annulus protects the scintillator counter annulus from muons, so that it can be used to detect and reject positrons 
from decay muons in the Ag annulus.}
\label{fig2}
\end{figure}

As in our previous study of the OP91 and OD80 Bi2212 samples, the ZF-$\mu$SR measurements reported here were performed on the M20D surface muon beam line 
at TRIUMF in Vancouver, Canada using the LAMPF Helmholtz-coil spectrometer 
and a muon veto cryostat insert. The latter utilizes a positive muon ($\mu^+$) plastic scintillator counter placed in front of the sample and a ``veto'' 
plastic scintillator detector positioned behind the sample to discard events from muons that miss or pass through the sample (see Fig.~\ref{fig2}). 
Both of these counters are contained within a helium-gas flow cryostat. The ZF-$\mu$SR signal is constructed from the detection of decay 
positrons originating from muons that trigger the internal $\mu^+$ counter, but not the veto detector. 

Nearly 100~\% spin-polarized $\mu^+$ 
were implanted in the sample with the initial muon-spin polarization ${\bf P}(t \! = \! 0)$ parallel to the muon beam (denoted as the $z$-direction) and 
the crystallographic $c$-axis of the sample. The external magnetic field at the sample position was nullified using three-axis Helmholtz coils.
To stop a sufficient number of muons in the OD70 single crystal, the muon beam momentum was reduced by 2.5~\% and a layer of 0.025~mm thick 99.998~\% pure silver (Ag) 
foil was placed in front of the sample to act as an energy degrader. We note that the muon beam momentum was reduced by 4~\% for our previous measurements 
on the OP91 and OD80 single crystals,\cite{Pal:18} but Ag foil was used as an energy degrader only for the measurements
on the OP91 sample. No reduction of the muon beam momentum or the use of Ag-foil as an energy degrader was necessary for the thicker OD55 single 
crystal. For comparison, we also performed ZF-$\mu$SR measurements on stacked layers of the high purity Ag foil with a total thickness of 0.25~mm. 
Silver is regarded as a reference material due to the very small size of its
nuclear moments, minimizing any muon spin depolarization processes. Table~\ref{Samples} lists the new samples and those previously studied,\cite{Pal:18}
and also indicates which samples were measured with the use of Ag degrader and/or a reduction of the muon beam momentum from the nominal value 
$p_{\mu} \! \simeq \! 29.3$~MeV/c.

\begin{table}
\caption{\label{Samples} Summary of samples and $\mu$SR measurement conditions.} 
\begin{ruledtabular}
\begin{tabular}{ccccc}
Sample &
Area & 
Area density\footnotemark[1] &
$\Delta \! p_{\mu}/p_{\mu}$\footnotemark[2] &
Degrader thickness\footnotemark[3] \\
       & (cm$^2$) & (mg/cm$^2$) &   & (mm) \\
\hline
OP91\footnotemark[4] & 0.25 & 140 & -4.0 $\%$ & 0.025 \\
OD80\footnotemark[4] & 0.39 & 185& -4.0 $\%$ & - \\
OD70 & 0.25  & 104& -2.5 $\%$ & 0.025 \\
OD55 & 0.68  & 172& 0 & - \\
OD80\footnotemark[5] & 0.18 & 216 & 0 & - \\
Ag & 0.56 & 262 & -2.5 $\%$ & - \\
Ag\footnotemark[6] & 0.36 & 262 & 0 & - \\
Ag\footnotemark[5] & 0.18 & 262 & 0 & - \\
\end{tabular}
\end{ruledtabular}
\footnotetext[1]{Thickness of Ag needed to stop muons with momentum $p_{\mu} \! \simeq \! 29.3$~MeV/c is $\sim \! 0.19$~mm, 
corresponding to an area density of 199~mg/cm$^2$.}
\footnotetext[2]{Nominal muon beam momentum is $p_{\mu} \! \simeq \! 29.3$~MeV/c.}
\footnotetext[3]{99.998~\% pure Ag foil.} 
\footnotetext[4]{Stacked and tiled mosaic of single crystals studied in Ref.~\onlinecite{Pal:18}.}
\footnotetext[5]{Measurements carried out using an ultra-low background cryostat insert.}
\footnotetext[6]{Studied in Ref.~\onlinecite{Pal:18}.}  
\end{table} 

\section{Data Analysis and Results}
\begin{figure}
\centering
\includegraphics[width=\columnwidth]{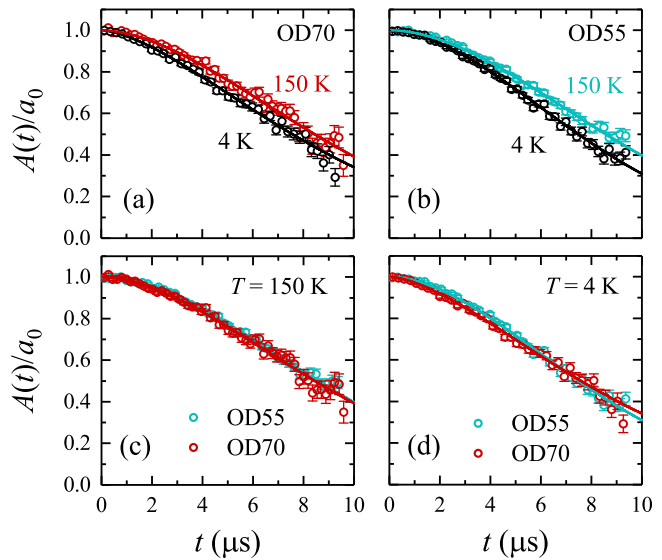}
\caption{Normalized ZF-$\mu$SR asymmetry spectra for (a) OD70 at $T \! = \! 150$~K (red circles) and $T \! = \! 4$~K (black circles), and (b) OD55 at $T \! = \! 150$~K (blue circles) and $T \! = \! 4$~K (black circles). Normalized ZF-$\mu$SR asymmetry spectra for OD70 (red circles) and OD55 (blue circles) at (c) $T \! = \! 150$~K, and (d) $T \! = \! 4$~K. The solid curves are fits to $A(t) \! = \! a_0 G_z(t)$, where $G_z(t)$ is given by Eq.~(\ref{Eq:stretched}).}
\label{fig3}
\end{figure}

\begin{figure}
\centering
\includegraphics[width=\columnwidth]{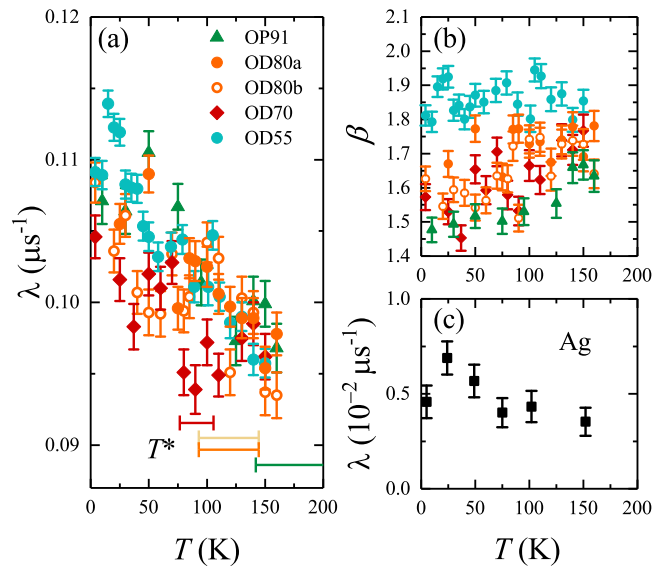}
\caption{Temperature dependence of (a) the ZF relaxation rate $\lambda$ and (b) the stretching exponent $\beta$ 
resulting from fits of the ZF-$\mu$SR asymmetry spectra assuming Eq.~(\ref{Eq:stretched}). 
The two data sets for the OD80 sample in (a) correspond to data collected during two separate experiments, demonstrating 
the reproducibility of the results. The range of $T^*$ values shown at the bottom of panel (a) are compiled values from angle-resolved photoemission 
spectroscopy and tunneling experiments.\cite{Vishik:12}  
(c) Temperature dependence of the ZF relaxation rate in high purity Ag measured in the same experimental setup. Note that $\lambda$ in this case comes 
from fits to Eq.~(\ref{Eq:stretched}) with $\beta \! = \! 1$.}
\label{fig4}
\end{figure}

Figure~\ref{fig3} shows representative ZF-$\mu$SR spectra for the OD55 and OD70 single crystals. While Figs.~\ref{fig3}(a) and \ref{fig3}(b) indicate a 
slightly enhanced relaxation rate with decreasing temperature for both samples, Figs.~\ref{fig3}(c) and \ref{fig3}(d) show that the temperature dependence 
does not evolve with hole doping. The ZF spectra are described by the equation
\begin{equation} 
A(t) \! = \! a_0 G_z(t) \, 
\end{equation}  
where $a_0$ is the initial asymmetry and $G_z(t)$ is a ZF relaxation function that describes the time evolution of the muon spin polarization $P(t)$ 
due to internal magnetic fields sensed by the muon ensemble inside the sample.
The following ZF relaxation function is commonly assumed for high-$T_c$ cuprate superconductors
\begin{equation}
G_z(t)= G_{\mathrm{KT}}(\Delta,t)\mathrm{exp}(-\lambda t),
\label{1component}
\end{equation}
where $G_{\mathrm{KT}}(\Delta,t)$ is a temperature-independent static Gaussian Kubo-Toyabe (GKT) function. The GKT function approximates relaxation of $P(t)$ by a dense 
system of randomly oriented nuclear dipole moments, which on the $\mu$SR time scale generate an internal static magnetic field distribution of width $\Delta/\gamma_\mu$, where
$\gamma_\mu$ is the muon gyromagnetic ratio. 
The exponential relaxation function is intended to account for any temperature-dependent relaxation caused by electronic moments.
Since it has been shown that the GKT function does not adequately describe the relaxation of the ZF-$\mu$SR signal in La214 by the
nuclear moments,\cite{Huang:12} here we make no assumptions about the functional form of the relaxation functions that describe the nuclear and electronic moment
contributions, and instead fit the ZF-$\mu$SR asymmetry spectra for Bi2212 to a phenomenological stretched-exponential relaxation function of the form
\begin{equation}
G_z(t)=\mathrm{exp}\left[-(\lambda t)^\beta\right]\ \, .
\label{Eq:stretched}
\end{equation}
Any magnetism associated with the PG phase should evolve with changes in hole-doping concentration, and if fluctuating slow enough to affect the muon spin relaxation,
will cause a corresponding change in the ZF relaxation rate. To determine whether there is any doping dependence, we carried out global fits of the 
ZF-$\mu$SR spectra for the OD55 and OD70 samples, as well as the previously studied OP91 and OD80 samples.\cite{Pal:18}
Both $\lambda$ and $\beta$ were treated as temperature dependent fitting parameters. Representative fits are shown in Fig.~\ref{fig3} and the fitting results
are shown in Figs.~\ref{fig4}(a) and \ref{fig4}(b). Typically, the fitting parameters $\lambda$ and $\beta$ are expected to be independent of temperature if 
the relaxation of the ZF signal is solely due to nuclear magnetic moments. Although this is clearly not the case, the increase in $\lambda$ with decreasing 
temperature is similar for all of the Bi2212 samples. We note that $\beta$ is larger and independent of temperature for the OD55 single crystal, which was
measured with nominal beam momentum and no Ag degrader foil. 

Since a PG is not expected in the OD55 sample, except perhaps in a $p \! < \! 0.22$ minority phase occupying a small 
volume fraction, the temperature-dependent ZF relaxation rate cannot be associated with the PG phase. This raises the question of whether the source 
is an intrinsic property of Bi2122 or a background contribution from muons stopping outside of the sample.     

\section{Ultra-Low Background Measurements}
The standard muon veto cryostat insert used for experiments here, in Ref.~\onlinecite{Pal:18}, and for a number of the other previous ZF-$\mu$SR studies 
of high-$T_c$ cuprates (including Ref.~\onlinecite{Zhang:18}), is designed for samples with an area of $\sim \! 0.25$~cm$^2$ to $9.6$~cm$^2$. 
The size of some of the Bi2212 samples listed in Table~\ref{Samples} are at or near the lower limit. To determine if a background contribution 
from muons passing through or missing the sample is the source of the temperature-dependent ZF relaxation rate, we repeated measurements on one of the OD80 
single crystals of Ref.~\onlinecite{Pal:18} using an ``ultra-low background'' (ULB) cryostat insert specifically designed for small samples.
The OD80 single crystal has an area of 0.18~cm$^2$ and mass of $\sim \! 39$~mg, and is thick enough to negate the need for placing Ag foil 
in front of the sample or reducing the muon beam momentum. 
Here we note that the muon beam momentum was reduced by 4~\% for the measurements on mosaics of OD80 and OD91 single crystals reported 
in Ref.~\onlinecite{Pal:18}.  
The ULB insert achieves a negligible background signal by the placement of a second kind of internal veto counter (``active collimator'') 
upstream of the sample, as described in the caption of Fig.~\ref{fig2}. The trade-off is a significantly slower count rate compared to the standard 
muon veto cryostat insert.

\begin{figure}
\centering
\includegraphics[width=\columnwidth]{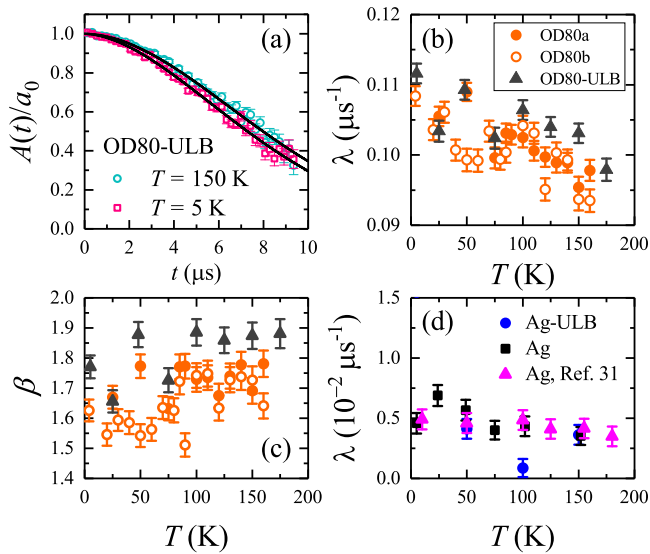}
\caption{(a) Normalized ZF-$\mu$SR asymmetry spectra from ULB measurements of the OD80 Bi2212 sample at $T \! = \! 100$~K (blue circles) and $T \! = \! 5$~K (red squares). Solid 
curves are fits to Eq.~(\ref{Eq:stretched}). Temperature dependence of (b) the ZF relaxation rate and (c) $\beta$ for the OD80 sample 
from measurements using the ULB cyrostat insert (grey triangles) and two sets of measurements in different muon 
beam periods using the standard muon veto cryostat insert (orange circles). (d) Temperature 
dependence of the ZF relaxation rate in high purity Ag from experiments using the ULB (blue circles) and the standard (black squares) muon veto cryostat inserts,
and previous results\cite{Pal:18} (pink triangles) using the standard muon veto cryostat insert. The values of $\lambda$ for Ag are from fits assuming $\beta \! = \! 1$.}
\label{fig5}
\end{figure}

Representative ZF-$\mu$SR asymmetry spectra from the ULB measurements on the OD80 single crystal are shown in Fig.~\ref{fig5}(a) 
together with fits assuming the relaxation function of Eq.~(\ref{Eq:stretched}). Figures~\ref{fig5}(b) and \ref{fig5}(c)
show the ensuing temperature dependences of the relaxation rate $\lambda$ and $\beta$ compared to the results obtained using
the standard muon veto cryostat insert. For the most part, the ULB measurements of OD80 produce larger values of $\lambda$ and $\beta$ 
that are comparable to the values obtained from measurements of the large OD55 single crystal using the standard muon cryostat insert
without reduced beam momentum or Ag degrader (see Table~\ref{Samples}). For comparison, Fig.~\ref{fig5}(d) shows that the relaxation
rate of 0.25~mm-thick Ag foil samples measured using the standard and ULB cryostat inserts are similar. 
These results and an inspection of the sample sizes and measurement conditions in Table~\ref{Samples} indicate that the unsystematic variation 
of $\lambda$ and $\beta$ with hole-doping in Figs.~\ref{fig4}(b) and \ref{fig4}(c) is due to muons stopping outside the sample
in the standard muon cryostat insert setup. In particular, at lower beam momentum a greater fraction of muons are scattered at wide angles
upstream of the sample and the veto detection efficiency is reduced by the lower energy muons.    
There is also a temperature-independent contribution from muons stopping in the Ag-degrader foil placed in front of some of 
the Bi2212 samples. Irrespective of the various background contributions, the OD80 results obtained with the ULB setup show that the 
temperature dependence of $\lambda$ is predominantly an intrinsic property of Bi2212.       

\begin{figure*}
\centering
\includegraphics[width=\textwidth]{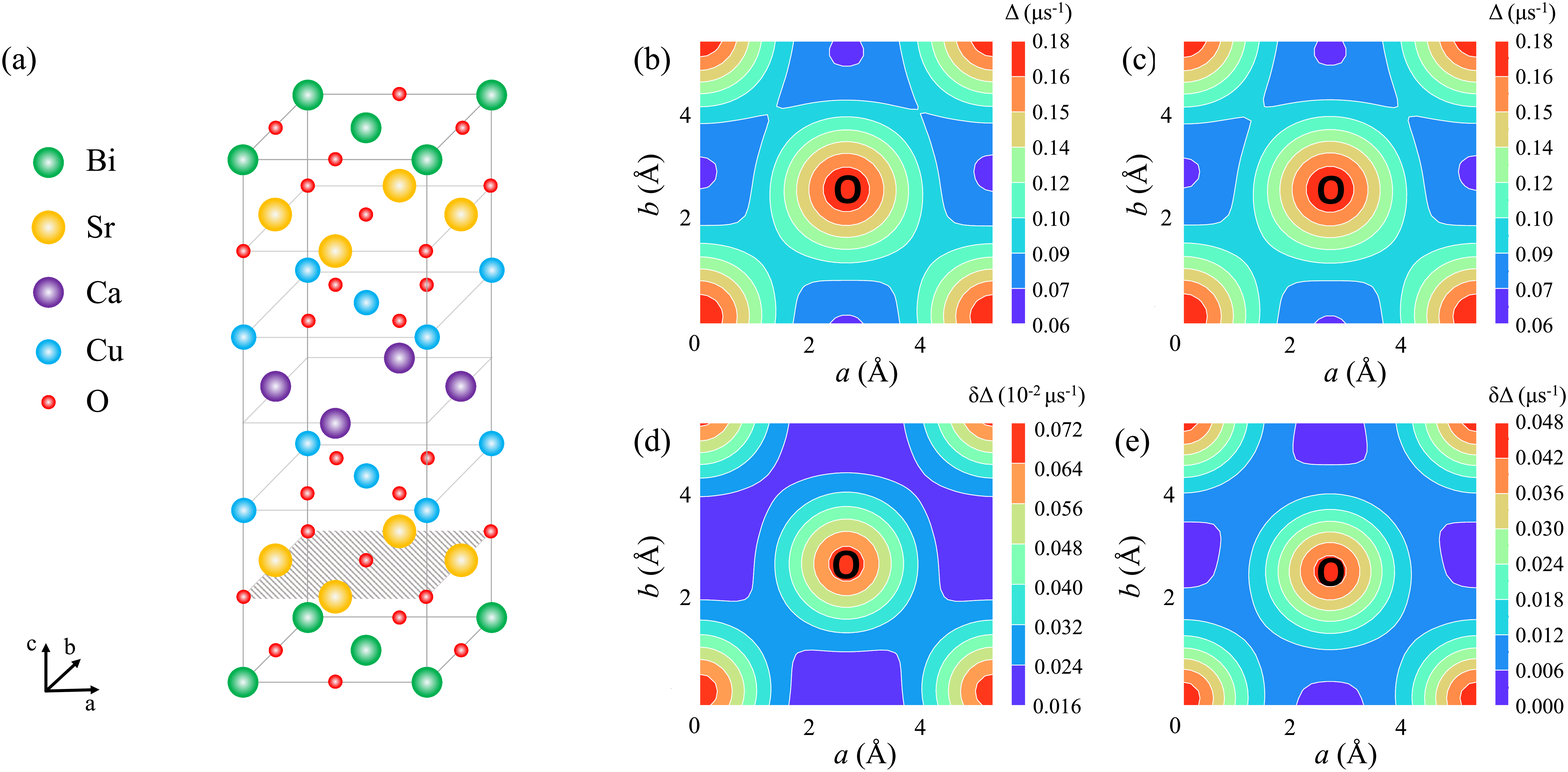}
\caption{(a) The orthorhombic crystal structure of Bi2212 (only half of the crystal structure is illustrated for clarity). Contour plots of $\Delta$ in the SrO layer (shaded plane) are calculated for 
(b) $T \! = \! 150$~K and (c) $T \! = \! 5$~K. (d) The difference between the $\Delta$ values in (b) and (c) at $100 \! \times$ magnification. 
(e) The difference between the $\Delta$ values within the SrO layer for $T \! = \! 150$~K and within a parallel plane
0.15~\AA~ toward the nearest BiO layer for $T \! = \! 5$~K. Note that the large values for $\Delta$ near the apical oxygen atoms 
are due to the Cu and Bi nuclei located directly below or above in the adjacent CuO$_2$ and BiO layers.}
\label{fig6}
\end{figure*}

\section{Nuclear Dipole Fields}
The crystal structure of Bi2212 is orthorhombic.
Thermal expansion,\cite{Yang:95} internal friction,\cite{Anderson:97} and neutron diffraction\cite{Miles:97} measurements on Bi2212 show that the
lattice parameters decrease with decreasing temperature and a subtle structural change occurs near 160~K. While the shrinking of the $a$-axis and $c$-axis
lattice parameters ceases near $\sim \! 50$~K and $\sim \! 90$~K, respectively, the contraction of the $b$-axis lattice parameter persists to much lower temperature. In general,
structural changes modify the nuclear dipole contribution to the ZF-$\mu$SR signal by changing the distance between the muon stopping site and the host nuclei,
and by changing the interaction of spin $I \! \neq \! 1/2$ nuclei that possess an electric quadrupole moment with the electric field gradient (EFG) at the nuclei. 
For example, a significant structural change in Y123 near 60~K has previously been argued to be the source of observed changes in the ZF-$\mu$SR 
relaxation rate.\cite{Sonier:02,Sonier:09}   

The magnitude ($\sim \! 1$~G) and quasistatic nature of the magnetic fields detected in Bi2212 by ZF-$\mu$SR\cite{Pal:18} are characteristic of nuclear dipole fields. 
To understand if changes in the lattice parameters of Bi2212 can be responsible for the observed change in the ZF relaxation rate with temperature, we have calculated the second moment of the
distribution of nuclear dipolar fields at various potential muon stopping sites. 
In Bi2212, the nuclei that contribute to the nuclear dipolar fields are $^{209}$Bi with spin $I \! = \! 9/2$ and magnetic moment $\mu_{\rm Bi} \! = \! 4.1103$~$\mu_{\rm N}$, 
and the two $I \! = \! 3/2$ isotopes $^{63}$Cu and $^{65}$Cu, which have a magnetic moment weighted by natural abundance of $\mu_{\rm Cu} \! = \! 2.275$~$\mu_{\rm N}$. 
The second moment of the corresponding Gaussian distribution of nuclear dipolar fields $B_{\rm dip \, N}$ for a particular muon site is given by
\begin{equation}
 \langle (B_{\rm dip \, N} - \langle B_{\rm dip \, N} \rangle)^2 \rangle = \frac{\Delta^2}{\gamma_\mu^2}=\frac{2}{3}\sum_{i,j} \mu_{i}\frac{1}{r^6_{i,j}} \, ,
\end{equation}
where $i$ indexes the nuclear species and $j$ indexes the location of the $j$th nucleus. 
The 2/3 prefactor is intended to account for the modification of the EFG at the nuclei
by the unscreened positive muon.\cite{Nishida:90,Hayano:79} The muon stopping sites in Bi2212 are unknown, but it is widely believed that in cuprates the positive muon forms a 
hydrogen-like bond with an oxygen atom. Previous studies show that the muon resides near an apical oxygen in La214,\cite{Huang:12} whereas in Y123 there are two muon stopping sites
--- one near an apical oxygen and the other near an oxygen atom in the CuO chain layer.\cite{Weber:91,Pinkpank:99} Since Bi2122 does not have CuO chains, we have calculated  
$\Delta$ for sites in the SrO layer where the apical oxygen reside, using values of the lattice parameters for $T \! = \! 150$~K and $T \! = \! 5$~K 
reported in Ref.~\onlinecite{Miles:97}. Figures~\ref{fig6}(a) and \ref{fig6}(b) show contour plots of the calculated $\Delta$ values, and Fig.~\ref{fig6}(c) shows the difference in these
values for the two temperatures, which we denote by $\delta \Delta$. 

Fits of the OD80 asymmetry spectra recorded using the ULB cryostat insert to the following asymmetry function
\begin{equation}
    A(t)=a_0 G_{\rm KT}(\Delta, t) \, ,
\label{eq:GKT}
\end{equation}
yield $\Delta \! = \! 0.1036(8)$~$\mu$s$^{-1}$ and 0.1136(8)~$\mu$s$^{-1}$ for $T \! = \! 150$~K and $T \! = \! 5$~K, respectively --- and hence a change
$\delta \Delta \! = \! 0.010(1)$~$\mu$s$^{-1}$ between these two temperatures. While there are no sites in the SrO layer where both the calculated values of
$\Delta$ and $\delta \Delta$ match the experimental values, the muon is expected to reside somewhat out
of the SrO layer. Moreover, the angle of the muon bond with the apical oxygen atom is expected to change slightly with the change in lattice parameters.  
Figure~\ref{fig6}(d) shows calculated values of $\delta \Delta$ assuming the muon site at $T \! = \! 5$~K is slightly displaced by 0.15~\AA~ out of the SrO layer 
towards the BiO layer. The region of radius $\sim \! 1.5$~\AA~ centered about the apical oxygen atom produces not only values for $\Delta$ that are in agreement with the fits
to Eq.~(\ref{eq:GKT}) ($\Delta \! \sim \! 0.1$~$\mu$s$^{-1}$), but also a change in $\Delta$ similar to what is measured ($\delta \Delta \! \sim \! 0.01$~$\mu$s$^{-1}$).
We note that there are no sites in or near the BiO and CuO$_2$ layers where the calculated values of $\Delta$ and $\delta \Delta$ agree with the experimental results. 
This does not mean that we have determined the muon site(s) in Bi2212, but rather have demonstrated that the observed temperature dependence of the ZF relaxation rate 
can be explained by changes in the nuclear dipole fields.

\section{Summary and Conclusions}
To summarize, we have shown that there is a weak temperature-dependent ZF-$\mu$SR relaxation rate in Bi2122 that does not evolve over a wide hole-doping range 
$0.16 \! \leq \! p \! \leq \! 0.23$ and persists outside the PG phase. Measurements using an ULB cryostat insert demonstrate that while the temperature dependence in the relaxation rates is an intrinsic property of Bi2212, the doping dependence in $\beta$ arises from muons stopping outside of the sample. Previous longitudinal-field $\mu$SR measurements showed that 
the relaxation is caused by quasistatic magnetic fields, which were presumed to be of electronic origin.\cite{Pal:18} Although the precise muon stopping site(s) in
Bi2212 are unknown, here we have shown that a change in the nuclear dipole contribution to the ZF-$\mu$SR signal associated with a small change in the lattice parameters
is the probable source of the temperature-dependent ZF-$\mu$SR relaxation rate. An evolution of the crystal structure with temperature should also influence
the nuclear dipole contribution in other high-$T_c$ cuprates, although the effect may be weaker or difficult to disentangle from other influences. Like Bi2212, the crystal structure
of Y123 evolves over a wide temperature range.\cite{You:91,Meingast:91} Yet different from Bi2212, the nuclear dipole contribution in Y123 presumably varies
with hole doping due to a change in the relative population of the muon sites\cite{Pinkpank:99} and charge order in the CuO chains.
Our results on Bi2212 show that it is generally not valid to assume the nuclear dipole contribution to the ZF-$\mu$SR asymmetry spectrum of
cuprate high-$T_c$ superconductors is temperature independent.  

\begin{acknowledgments}
We thank the technical staff of TRIUMF's Centre for Molecular and Materials Science for assistance. J.E.S. and S.R.D. acknowledge support from the 
Natural Sciences and Engineering Research Council of Canada. Work at Brookhaven National Laboratory was supported by the U.S. Department of Energy, 
the Office of Basic Energy Sciences, and the Division of Materials Sciences and Engineering under Contract No. DE-SC0012704.
\end{acknowledgments}

\end{document}